\documentclass[12pt]{book}
\usepackage{plenum}
\usepackage{epsfig}
\input{epsf.tex}
\input psfig
\newcommand{\beq}{\begin{equation}}
\newcommand{\eeq}{\end{equation}}
\newcommand{\bea}{\begin{eqnarray}}
\newcommand{\eea}{\end{eqnarray}}
\newcommand{\un}[1]{{\it #1}}

\newcommand{\real}{\relax{\rm I\kern-.18em R}}

\newcommand{\site}[1]{\refnote{\cite{#1}}}
\newcommand{\Ref}[1]{Eq.(\ref{#1})}
\newcommand{\dfrac}[2]{\frac{\displaystyle{#1}}{\displaystyle{#2}}}
\newcommand{\ii}{{\rm i}}
\newcommand{\ee}{{\rm e}}
\begin{document}
\chapter{THE DUAL SUPERCONDUCTOR PICTURE FOR CONFINEMENT}

\author{Adriano Di Giacomo}

\affiliation{\affnote{1} Dip. Fisica Universit\`a and INFN, \\
Piazza Torricelli 2, 56100 Pisa, Italy}
\vskip1in
%\vskip1.0in\par\noindent
%Lectures delivered at the Workshop on {\em Non perturbative aspects of QCD}\\
%\par\noindent
%\hskip0.6\textwidth
%Newton Institute, Cambridge UK.\par\noindent
%\hskip0.6\textwidth
% June 22 - July 4\quad 1997
%\vfill\eject
\section{Introduction}
At short distances QCD vacuum mimics the Fock space ground state of 
perturbation theory: deep inelastic scattering experiments, jet production 
in high energy reactions, QCD sum rules provide empirical evidence for that.

In fact colour is confined, and quarks and gluons never appear as free 
particles in asymptotic states: the ground state is very different from the
perturbative vacuum.Its exact structure is not known.
Many models have been attempted to give an approximate description of it: some
of them are based on ``mechanisms'', i.e. they assume that 
the degrees of freedom relevant for large distance physics behave as other well
understood physical systems.

The most attractive mechanism for colour confinement is 
{\em Dual superconductivity of type II of QCD vacuum}\site{1,2}
%DUAL SUPERCONDUCTIVITY
%OF TYPE II OF QCD VACUUM. 
Dual means interchange of electric with magnetic with 
respect to ordinary superconductors. The idea is that the chromoelectric field 
in the region of space between a $Q\bar Q$ pair is constrained by dual Meissner
 effect into Abrikosov\site{3} 
flux tubes, with constant energy per unit length. The 
energy is then proportional to the distance
\begin{equation}
E = \sigma R\label{eq:1.1}
\end{equation}
and this means confinement.

Lattice is the ideal tool to study (at least numerically) large distance 
phenomena from first principles. There is indeed evidence from lattice
simulations that:
\begin{itemize}
\item[1)] The string tension exists: large Wilson loops $W(R,T)$ describing a 
pair of static quarks at a distance $R$ propagating for a time $T$, obey the
area law\site{4}
\begin{equation}
W(R,T) \simeq \exp(- \sigma R T)\label{eq:1.2}\end{equation}
Since in general $W(R,T) \simeq \exp(- V(R) T)$, the observed behaviour
\Ref{eq:1.2} confirms confinement,as defined by \Ref{eq:1.1}.
\item[2)] Chromoelectric flux tubes have been observed, joining $Q$ $\bar Q$ 
pair propagating in Wilson loops\site{5,5a}. 
Their transverse size is $\sim 0.5$~fm.
\item[3)] String like modes of these flux tubes have been detected\site{6}.
\item[4)] Particles belonging to higher representations than quarks also experience a string tension at intermediate distances, which, for $SU(2)$, 
depends on the colour spin $J$ as\site{7}
\[
\sigma_J = k J(J+1)
\]
or
\begin{equation}
\dfrac{\sigma_J}{\sigma_{1/2}} = \frac{3}{4} J (J+1)
\label{eq:1.3}
\end{equation}
\end{itemize}
Observations 1)-3) support the idea of dual superconductivity. We will discuss 
in detail the implications of 4) in the following.

The problem we adress in these lectures is: can dual supeconductivity
of QCD vacuum be directly tested?

The plan of the lectures is as follows. We will recall basic superconductivity
and its order parameter, in order to clarify what we are looking for.
We will then define dual superconductivity and its disorder parameter.
We will construct the disorder parameter for  $U(1)$ pure gauge theory.

We will then
check the construction with the $X-Y$ 3d model (liquid He4).

We will then revisit the abelian projection, which reduces the problem of dual 
superconductivity in QCD to a $U(1)$ problem. The Heisenberg ferromagnet will 
prove 
a useful laboratory to check this procedure. We will finally show how dual 
superconductivity can be directly detected in $SU(2)$ and $SU(3)$ 
gauge theories.

A
discussion of the results and of their physical 
consequences is contained in the final section.

\section{Basic superconductivity: the order parameter$^9$.}
A relativistic version of a superconductor is the abelian Higgs model
\beq
{\cal L} = -\frac{1}{4}F_{\mu\nu}F_{\mu\nu} +
( D_\mu\Phi)^\dagger (D_\mu\Phi) - V(\Phi)\label{eq:2.1}
\eeq
$F_{\mu\nu}$ is the electromagnetic field strength, $D_\mu$ is the covariant 
derivative
\beq
D_\mu\Phi = (\partial_\mu - i q A_\mu)\Phi\label{eq:2.2}
\eeq
and $V(\Phi)$ the potential of the scalar field
\beq
V(\Phi) = \frac{1}{4}\left(\Phi^\dagger\Phi - \mu^2\right)^2\label{eq:2.3}
\eeq
If $\mu^2$ is positive the field $\Phi$ has a nonzero vacuum expectation value.
Since $\Phi$ is a charged field this is nothing but a spontaneous breaking of 
the $U(1)$ symmetry related to charge conservation. The ground state is a 
superposition of states with different electric charges, a phenomenon which 
is usually called ``condensation'' of charges.

A convenient parametrization of $\Phi$ is
\beq
\Phi = \rho\ee^{\ii \theta q}\qquad \rho = \rho^\dagger > 0
\label{eq:2.4}
\eeq
Under gauge transformations
\beq
A_\mu\to A_\mu - \partial_\mu\alpha\qquad
\theta\to\theta+\alpha
\label{eq:2.5}
\eeq
The covariant derivative (\ref{eq:2.2}) reads in this notation
\beq
D_\mu\Phi = \ee^{\ii \theta q}\left[\partial_\mu - \ii q( A_\mu -
\partial_\mu\theta)\right]\rho\label{eq:2.6}
\eeq
The quantity $\tilde A_\mu = A_\mu - \partial_\mu\theta$ is gauge invariant. 
Moreover 
\beq
F_{\mu\nu} =\partial_\mu A_\nu - \partial_\nu A_\mu
=\partial_\mu \tilde A_\nu - \partial_\nu \tilde A_\mu\label{eq:2.6a}
\eeq
The equation of motion reads, neglecting loop corrections, 
(or looking at ${\cal L}$ as an effective lagrangean)
\beq
\partial_\mu F^{\mu\nu} + \dfrac{\tilde m^2}{2} \tilde A_\nu = 0
\qquad
\tilde m= \sqrt{2} q \langle \Phi\rangle
\label{eq:2.7}\eeq
In the gauge $A^0=0$ a static configuration has $\partial_0 \vec A = 0$, 
$\partial_0\Phi = 0$ so that $E_i= F_{0i} = 0$. \Ref{eq:2.7} implies that
\beq \vec\nabla\wedge\vec H +  \dfrac{\tilde m^2}{2} \vec{\tilde A} = 0
\label{eq:2.8}\eeq
The term ${\tilde m^2}/{2}\, \vec{\tilde A}$ in \Ref{eq:2.8} is a consequence 
of spontaneous symmetry breaking and is an electric stationary current (London
 current). A persistent current with $\vec E = 0$, means $\rho = 0$ since
$\rho\vec j= \vec E$ and hence superconductivity.

The curl of \Ref{eq:2.8}, reads
\beq \nabla^2\vec H - \dfrac{\tilde m^2}{2} \vec H = 0
\label{eq:2.9}\eeq
The magnetic field has a finite penetration depth $1/\tilde m$, 
and this is nothing but
 Meissner effect. The key parameter is $\langle\Phi\rangle$, which
is the order parameter for superconductivity: it signals spontaneous breaking  
of charge conservation. 

Besides $\lambda_A = 1/\tilde m$ there is another parameter with dimension of 
a length, $\lambda_\Phi = \mu^{-1}$. 

If $\lambda_A \geq \sqrt{2}\lambda_\Phi$ the superconductor is called IInd kind
otherwise it is Ist kind.

For a superconductor of first kind,there is Meissner effect for external magnetic field $H< H_c$ (critical field), for $H > H_c$ the field penetrates the bulk and superconductivity is destroyed. For second kind instead a penetration by Abrikosov flux tubes of transverse size $1/\tilde m$ is energetically favoured.
Flux tubes repel each other. By increasing the external field 
the number of
flux tubes increases.
When they touch  each other 
the field penetrates the bulk and
superconductivity is destroyed. 

In a dual
superconductor the role of electric and magnetic field is interchanged. 
The $U(1)$ symmetry
related to magnetic charge conservation is spontaneously broken, i.e. 
monopoles condense in the vacuum. An order parameter for dual 
superconductivity will then be the vacuum expectation value of a field 
carrying non zero magnetic charge.
\subsection{Monopoles.} The equations of motion for the electromagnetic field 
in the presence of an electric 
current $j_\mu$ and of a magnetic current $j_\mu^M$ 
are
\begin{eqnarray}
\partial^\mu F_{\mu\nu} &=& j_\nu\nonumber\\
\partial^\mu F^*_{\mu\nu} &=& j^M_\nu \label{eq:2.10}
\end{eqnarray}
If both $j_\mu$ and $j^M_\mu$ are zero (no charges, no monopoles) photons 
are free, and the equations of motion are invariant under the transformation
\begin{eqnarray}
F_{\mu\nu} &\to& \cos\theta\,F_{\mu\nu} + \sin\theta\, F^*_{\mu\nu}
\nonumber\\
F^*_{\mu\nu} &\to& \cos\theta\,F^*_{\mu\nu} - \sin\theta\, F_{\mu\nu}
\label{eq:2.11}\end{eqnarray}
for any $\theta$. In particular if $\theta = \pi/2$ 
%\Ref{eq:2.11} 
Eq.'s(15)
give
$\vec E\to \vec H$, $\vec H\to - \vec E$ which is known as duality 
transformation. In nature $j_\mu^M = 0$. The general solution of \Ref{eq:2.10} 
is then written in terms of vector potential $A_\mu$
\beq F_{\mu\nu} = \partial_\mu A_\nu - \partial_\nu A_\mu
\label{eq:2.12}\eeq
and 
$\partial^\mu F^*_{\mu\nu} =0$ is identically satisfied (Bianchi identities).

If a monopole exists Bianchi identities are violated. However they can be 
preserved, and with them the description in terms of $A_\mu$, by considering 
the 
monopole as the end point of a  thin solenoid (Dirac string) connecting 
it to infinity: the flux of the Coulomb like magnetic field, 
$\vec H = \frac{M}{4\pi}
\frac{\vec r}{r^3}$, $\Phi(H) = M$ is conveyed to infinity by the 
string\site{9}.

The string is invisible if the parallel transport of any electric charge around it is trivial:
\[
\ee^{\ii q \oint \vec A {\rm d}\vec x} =\ee^{\ii q \Phi(H)}
= \ee^{\ii 2\pi n} = 1
\]
or
\beq q M = 2\pi n\label{eq:2.13}\eeq
\Ref{eq:2.13} is known as Dirac quantization condition, and constrains 
the $U(1)$
group to be compact. If one insists to describe the system in terms of $A_\mu$
monopoles are non local objects with non trivial topology. One could introduce
dual vector potential $B_\mu$, such that $F^*_{\mu\nu} =
\partial_\mu B_\nu - \partial_\nu B_\mu$. Dual Bianchi identities 
would read $\partial^\mu F_{\mu\nu} = 0$, 
monopoles would be pointlike but
 electric charges could only 
exist  if dual strings were attached to them.

There is another acceptation of duality, which originates from 
statistical mechanics.
The prototype example is the 2d Ising model. The model is defined on a square 
lattice, by associating to each site $i$ a field $\sigma(i)$ wich can assume 
2 values, say $\pm 1$. The action can be written
\beq {\cal L} = - \beta \sum_{i,j}\sigma(i)\sigma(j)
\label{eq:3.1}\eeq
the sum running on nearest neighbours.
The partition function $K[\beta]$ is known exactly in the thermodinamical 
limit. At high $\beta$ (low temperatures) the system is magnetized 
$\langle\sigma(i)\rangle \neq 0$; at low $\beta$ it is disordered. A dual 
description can be given of the same system, by associating to each link 
(dual lattice site) a variable $\sigma^*$ with value $-1$ if the values of 
$\sigma$ in the sites connected by the link are the  same, $+1$ otherwise.

It can be rigorously proven that the partition function in terms of the new 
variables has the same form as the original one $K[\beta]$ 
\beq K^*[\beta] = K[\beta^*]\label{eq:3.2}\eeq
with the only change 
\beq 
\beta\to \beta^*\qquad
\beta^* = 
\frac{1}{2}{\rm arcsinh}\left(\dfrac{1}{{\rm sinh} 2\beta}\right)
\simeq
\frac{1}{\beta}\label{eq:3.3}\eeq
A relation like \Ref{eq:3.2} is called a duality relation. It maps high 
temperature (strong couplings) regimes of $K^*$ to the low 
temperature (weak coupling) of $K$\site{10}.

Similar relations have been recently discovered in SUSY QCD with $N=2$, and more generally in models of string theory\site{11}.
\vskip0.3in
\par\noindent
%{\centerline
\hskip0.05\textwidth
\begin{minipage}{0.90\textwidth}
\epsfxsize = 0.8\textwidth
%{\centerline{
\epsfbox{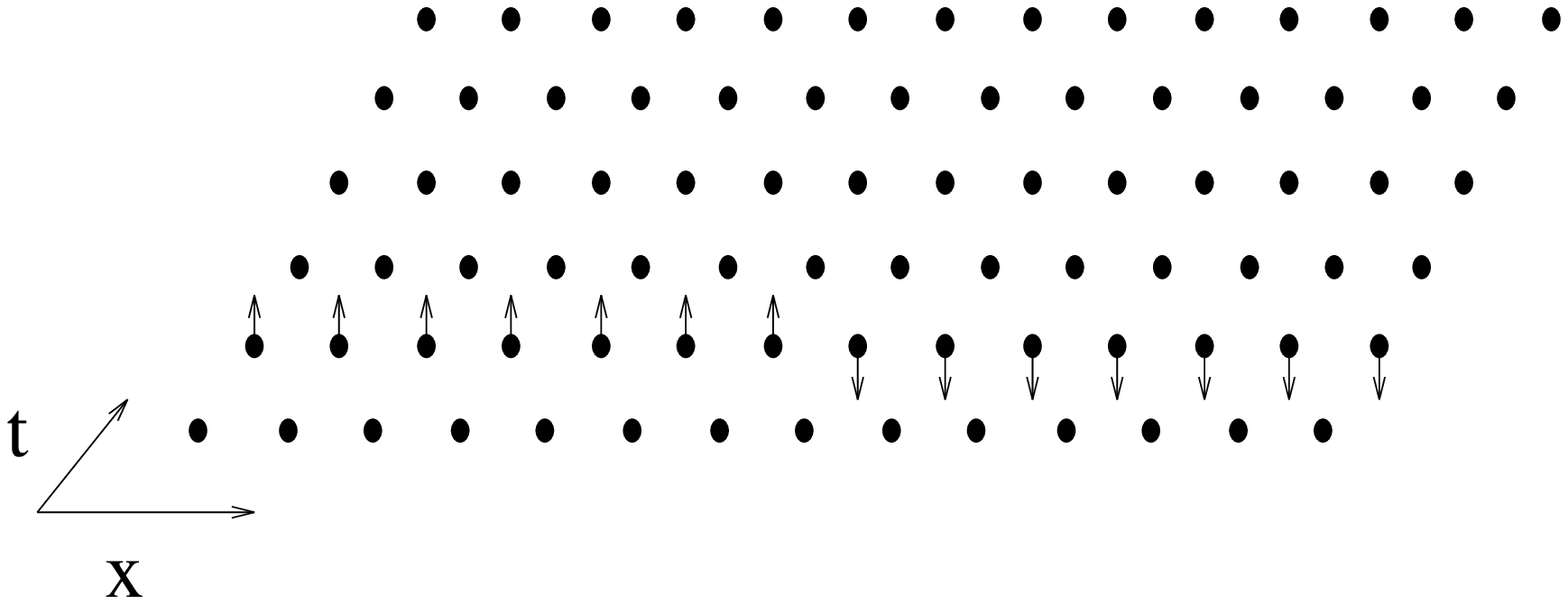}
% }}
{\centerline{
Fig.1 
 A kink in Ising model.
}}
\end{minipage}
%}
\vskip0.1in\par\noindent
If we look at the Ising model as the euclidean version of some 1+1 
dimensional field theory, a configuration at fixed $t$, with 
$\sigma(i) = -1$, $i < i_0$
$\sigma(i) = +1$, $i \geq i_0$ 
will appear on the dual lattice as a single spin 
up. This configuration has topology, it is a kink. The exitations of the dual
lattice are kinks. At low temperature the system is magnetized 
$\langle\sigma\rangle\neq 0$ and very few kinks are present. 
At high temperature 
$\langle\sigma\rangle =  0$, but, by duality relation, 
$\langle\sigma^*\rangle\neq 0$. $\langle\sigma^*\rangle$ is called a disorder parameter, as opposite to $\langle\sigma\rangle$ which is the order parameter.
The relation
\beq
\langle\sigma\rangle\cdot\langle\sigma^*\rangle = 0
\label{eq:3.3a}\eeq
can be proven in the thermodynamical limit. 
$\langle \sigma^*\rangle$ signals the condensation of kinks.

We shall next adress the study of monopole condensation in $U(1)$  compact 
gauge theory. We shall define a disorder parameter for this system, which 
describes the condensation of monopoles in the vacuum at high temperature 
(low $\beta$).
The parameter will be the v.e.v. of an operator with non zero monopole charge, 
and will thus signal dual superconductivity.

The same construction will then be used for non abelian gauge theories after abelian projection.
\section{Monopole condensation in compact $U(1)$: a disorder 
parameter$^{13}$.}
Like any other gauge theory, compact $U(1)$ is defined in terms of 
parallel transport along the links joining nearest neighbours on the lattice
\beq U_\mu(n) = \exp(\ii e a A_\mu(n))\label{eq:4.1}\eeq
$a$ being the lattice spacing.

In the following we shall denote $e a A_\mu(n)$ as $\theta_\mu(n)$. The action
is written in terms of the parallel transport $\Pi_{\mu\nu}$ around 
the elementary square of the lattice in the plane $\mu,\nu$
\begin{eqnarray}
\Pi_{\mu\nu} &=&
\exp\ii\left[\theta_\mu(n) + \theta_\nu(n+\hat\mu) - \theta_\mu(n+\hat\mu)
-\theta_\nu(n)\right]
\equiv \exp(-\ii \theta_{\mu\nu}(n))\nonumber\\
 \theta_{\mu\nu}(n)) &=& \Delta_\mu \theta_\nu(n) - \Delta_\nu\theta_\mu(n)
\mathop\simeq_{a\to 0} a^2 e F_{\mu\nu}\label{eq:4.2}
\end{eqnarray}
\beq
S =  \beta\sum_{n,\mu<\nu} (1 - \cos\theta_{\mu\nu})\label{eq:4.3}\eeq
As $a\to 0$
\beq S\simeq \beta\sum_{n,\mu\neq \nu} \dfrac{\theta^2_{\mu\nu}(n)}{4}
\label{eq:4.2a}\eeq
and \Ref{eq:4.3} describes photons if $\beta = 1/e^2$.

The generating functional of the theory (partition function) is
\beq Z(\beta) = \int\prod_{n,\mu}\left[ \dfrac{d\,\theta_\mu(n)}{2\pi}\right]
\exp(-S)\label{eq:4.4}\eeq
The theory is compact, since $S$ depends on the $\cos$ of the angular variables, and is invariant under change of variables
\beq \theta_\mu(n) \to \theta_\mu(n) + f_\mu(n)
\label{eq:4.4a}\eeq
with arbitrary $f_\mu(n)$. A special case of \Ref{eq:4.4a} are gauge transformations.

A critical $\beta$, 
$\beta_c \simeq 1.011$ exists such that for 
$\beta > \beta_c$ the theory describes free photons. 
For $\beta < \beta_c$ electric charge is confined: 
Wilson loop obey area law\site{13} \Ref{eq:1.2} and flux tubes are 
observed\site{14}.

A variant of the theory is provided by the Villain action
\beq \exp(-S) =
\sum_m\exp\left[-\frac{\beta}{2}\sum_{n,\mu<\nu}\left|\theta_{\mu\nu} -
2\pi m\right|^2\right]\label{eq:4.5}\eeq
For this variant, condensation of monopoles has rigorously been proven 
as a mechanism of confinement\site{15}.
Recently the proof has been extended to more general forms of the action, 
including Wilson action \Ref{eq:4.2}\site{16}. Monopoles are identified and 
counted by the following procedure\site{17}.

Since by construction $\pi \leq \theta_\mu(n)\leq \pi$, 
it follows from \Ref{eq:4.2} that
\[ -4\pi \leq \theta_{\mu\nu}(n) \leq 4 \pi\]
$\theta_{\mu\nu}$ can be redefined modulo an integer 
multiple of $2\pi$, $n_{\mu\nu}$ as
\beq 
\theta_{\mu\nu}= \bar\theta_{\mu\nu} + 2\pi n_{\mu\nu}
\qquad -\pi <  \bar\theta_{\mu\nu} \leq \pi
\label{eq:4.6}\eeq
and the monopole current as
\beq\rho^M_\mu = \frac{1}{6}\varepsilon_{\mu\nu\rho\sigma}
\Delta_\nu n_{\rho\sigma} \equiv \partial_\nu F^*{\mu\nu}
\label{eq:4.7}\eeq
The total number of monopoles is
\beq N^M = \sum_n \rho_0^M(n)
\label{eq:4.8}\eeq
$N^M$ is large in the confined phase, and drops to zero in the 
deconfined phase. It has sometime been identified with the disorder parameter 
for monopole condensation. Of course $N^M$ commutes with the monopole charge, 
and therefore cannot signal by any means spontaneous breaking of magnetic 
$U(1)$.
\subsection{The disorder parameter.}
The basic idea of the construction\site{10,17a} 
of a disorder parameter
is the simple formula for translations
\[ \ee^{\ii p a} | x\rangle = |x + a\rangle\]
If we identify in our field theory
\begin{eqnarray}
x &\to & \vec A(\vec x,t)\nonumber\\
p &\to& \vec E(\vec x,t) =
-\ii\dfrac{\delta}{\delta\vec A(\vec x,t)}\label{eq:4.9}\end{eqnarray}
then the operator
\beq
\mu(\vec y,t) =
\exp\left[\frac{\ii}{e}\int d^3 x\,\vec E(\vec x,t)\vec b(\vec x - \vec y)
\right]\label{eq:4.9a}\eeq
operating on field states in the Schr\"odinger representation will give
\[
\mu | \vec A(\vec x,t)\rangle =
| \vec A(\vec x,t) + \frac{1}{e}\vec b(\vec x-\vec y)\rangle
\] 
i.e. it will add 
a monopole
to any field configuration  provided that
$\vec b$ 
is the vector potential describing the field produced by the monopole
\beq
\frac{1}{e}\vec b(\vec r) = \frac{1}{e}
\frac{m}{2\pi}\dfrac{\vec r\wedge \vec n}{r(r - \vec r \cdot\vec n)}
\label{eq:4.10}\eeq
The gauge has been chosen to have the Dirac string in the direction $\vec n$.

$\mu$ is independent 
on the choice of the gauge for $\vec b$
if $\vec E$ obeys Gauss law.

On a lattice\site{17b}, after Wick rotation, and with the identification
\beq
a^2 E^i = \frac{1}{e}{\rm Im} \Pi^{0i} + {\cal O}(a^4)\simeq\frac{1}{e}
\sin\theta^{0i}\label{eq:4.11}\eeq
\beq
\mu(\vec y,n_0) =\exp\left[-\beta\sum_{\vec n }
b^i(\vec n -\vec y)\sin\theta^{0i}(
\vec n,n_0)\right]\label{eq:4.12}\eeq
Here $ b_i(\vec n)$ is the discretized transcription of \Ref{eq:4.10} and the 
$\beta$ in front comes from the normalization $1/e$ in \Ref{eq:4.1} times the 
$1/e$ appearing in the monopole charge.

A better definition (compactified) which shifts the angle $\theta_{0i}$ 
and not $\sin\theta_{0i}$ is\site{11bis}
\beq
\mu(\vec y,n_0) =
\exp\beta\sum_n\left[ S(\theta^{0i}(\vec n,n_0) + 
b^i(\vec n-\vec y)) - S(\theta^{0i}(\vec n,n_0))\right]
\label{eq:4.13}\eeq
which reduces to \Ref{eq:4.12} at first order in $\vec b$.

In \Ref{eq:4.13} by $S$ we denote the density of action. The action can be 
any form, e.g. \Ref{eq:4.3} or (\ref{eq:4.5}) provided 
\Ref{eq:4.2a} 
is satisfied.

If any number of monopoles or antimonopoles are created at time $n_0$, 
then $b^i(\vec n - \vec y)$ has to be replaced by the corresponding field 
configuration. To compute correlation functions of operators at different 
times, the rule is
\beq
\langle\mu(\vec y_1,n^0_1),\ldots\mu(\vec y_k,n^0_k)\rangle
=
\frac{1}{Z}\int\prod\left[\dfrac{d \theta_\mu(n)}{2\pi}\right]
\exp\left(\beta(S + \Delta S)\right)
\label{eq:4.14}\eeq
where
$S + \Delta S$ is obtained by replacing in the action the plaquettes $\Pi^{0i}
(\vec n,n^0_a) = 1 - \cos(\theta^{0i}(\vec n,n^0_a))$ by
$1 - \cos(\theta^{0i}(\vec n,n^0_a)+b^i(\vec n-\vec y_a))$.
$(1 \leq a\leq k)$.

In particular we will study the correlator
\beq {\cal D}(x_0) = \langle \mu(\vec 0,x_0)\mu(\vec 0,0)\rangle
\label{eq:4.14a}\eeq
At large enough $x_0$
\beq  {\cal D}(x_0) \mathop\simeq_{|x_0|\to\infty}
A \exp(-M|x_0|) + \langle \mu\rangle^2
\label{eq:4.15}\eeq
The last equality follows from cluster property, translation invariance and 
$C$ invariance of the vacuum. Our aim will be to extract $M$ and 
$\langle\mu\rangle$ from numerical determinations of 
${\cal D}(x_0)$.
$\langle\mu\rangle$ is the disorder parameter: a non zero value of it in the thermodinamic limit signals dual superconductivity. $M$ is the mass of the lightest excitation with the quantum numbers of a monopole and is a lower limit to the mass of the effective Higgs field which produces superconductivity.

As explained above
\beq  {\cal D}(x_0) = \dfrac{1}{Z[S]} Z[S+\Delta S]
\label{eq:4.16}\eeq
where $S+\Delta S$ is obtained from $S$ by the change
\begin{eqnarray}
&&
\theta^{i0}(\vec n,0) \to
 \theta^{i0}(\vec n,0) + b^i(\vec n)\label{eq:4.17}\\
&&
\theta^{i0}(\vec n,x_0) \to
 \theta^{i0}(\vec n,x_0) - b^i(\vec n)\label{eq:4.18}
\end{eqnarray}
and
$ b^i(\vec n)$ is defined by \Ref{eq:4.10}. Since
\[  \theta^{i0}(\vec n,0) =
-\theta^i(\vec n,1) + \theta^i(\vec n,0) +
\theta^0(\vec n +\hat i,0) - \theta^0(\vec n,0)\]
the replacement
\Ref{eq:4.17} amounts to the change
\beq \theta^i(\vec n,1) \to \theta^i(\vec n,1) - b_i(\vec n) \equiv
\bar\theta^i(\vec n,1)\label{eq:4.17a}\eeq
The change \Ref{eq:4.17a} can be reabsorbed in a 
redefinition
of variables of the Feynman integral defining 
 $Z[S+\Delta S]$ which leaves the measure invariant [Eq.(27)]. As a consequence 
\[\theta^{ij}(\vec n,1) \to \theta^{ij}(\vec n,1) +
\Delta^i b^j - \Delta^j b^i\]
meaning that a monopole is added at $x_0 = 1$.
Moreover
\[\theta^{0i}(\vec n,1) \to \theta^{0i}(\vec n,1) +
b^i(\vec n)\]
Again this change can be reabsorbed by a change of variables
\[  \theta^i(\vec n,2) \to \theta^i(\vec n,2) - b_i(\vec n)\]
after which
\[\theta^{ij}(\vec n,2) \to \theta^{ij}(\vec n,2) +
\Delta^i b^j - \Delta^j b^i\]
and
\[ \theta^{i0}(\vec n,2) \to  \theta^{i0}(\vec n,2) + b^i(\vec n)\]
The construction can be repeated till $n_0 = x_0$, 
when the addition of $b_i(\vec n)$ 
to $\theta^{i0}(\vec n,n_0)$
cancels with the term $-b_i(\vec n)$ in 
\Ref{eq:4.18}. The change $Z[S + \Delta S]$ amounts to add a monopole 
in the site
 $\vec n = \vec y$, propagating from time 0 to time $x_0$.

Measuring ${\cal D}(x_0)$, to extract $M$ and $\langle\mu\rangle$ is non 
trivial, due to large fluctuations. $\Delta S$ is the change of the action on a spatial volume $V$: it fluctuates roughly as $\sqrt{V}$, which means a fluctuation $\sim\exp(\sqrt{V})\;$ on $\langle\mu\rangle$. A way out of this difficulty is to measure, instead of ${\cal D}(x_0)$, the quantity
\beq
\rho(x_0) = \dfrac{d}{d\beta}\ln\,{\cal D}(n_0) =
\langle S\rangle_S - \langle S+\Delta S\rangle_{S + \delta S}
\label{eq:4.19}\eeq
The last equality trivially follows from the definition of $Z$. 
$\langle S\rangle_S$ means 
average of $S$ computed by weighting with the action $S$.
Since ${\cal D}(x_0)_{\beta = 0} = 1$
\beq
{\cal D}(x_0) = \exp\left(\int_0^\beta\rho(\beta)d\,\beta\right)
\label{eq:4.20}\eeq
As $x_0\to\infty$ 
\beq \rho \mathop\simeq_{x_0\to \infty} 
2 \dfrac{d}{d\beta}\ln\langle\mu\rangle 
+ C \exp(-M x_0)\label{eq:4.21}\eeq
Fig.2 shows
a behaviour 
of $\rho$ as a function of $x_0$ consistent with 
\Ref{eq:4.21}.
\vskip0.1in
\par\noindent
\hskip0.05\textwidth
\begin{minipage}{0.90\textwidth}
\epsfxsize = 0.95\textwidth
\epsfbox{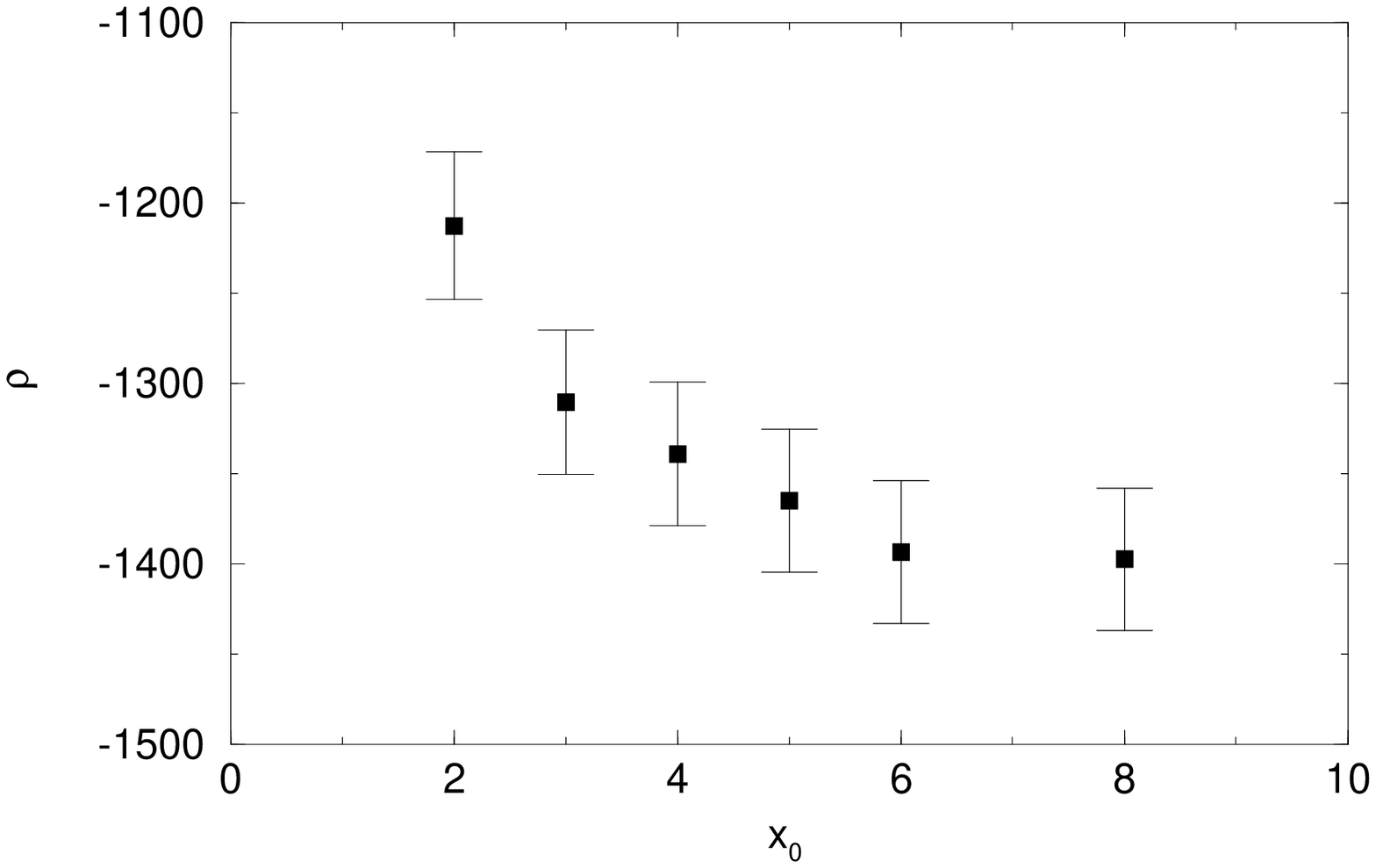}
Fig.2 Monopole antimonopole correlation.
(Lattice $8^3\times16$)
\end{minipage}
\vskip0.1in\noindent
Fig.3 shows $\rho(\infty) \equiv 2 \frac{d}{d\beta}\ln\langle\mu\rangle$
as a function of $\beta$. A huge negative peak appears at 
the phase transition, which, according to the definitions \Ref{eq:4.19}, 
\Ref{eq:4.20} reflects a sharp decrease of $\langle\mu\rangle$. 
\par\noindent
\hskip0.05\textwidth
\begin{minipage}{0.90\textwidth}
\epsfxsize = 0.95\textwidth
\epsfbox{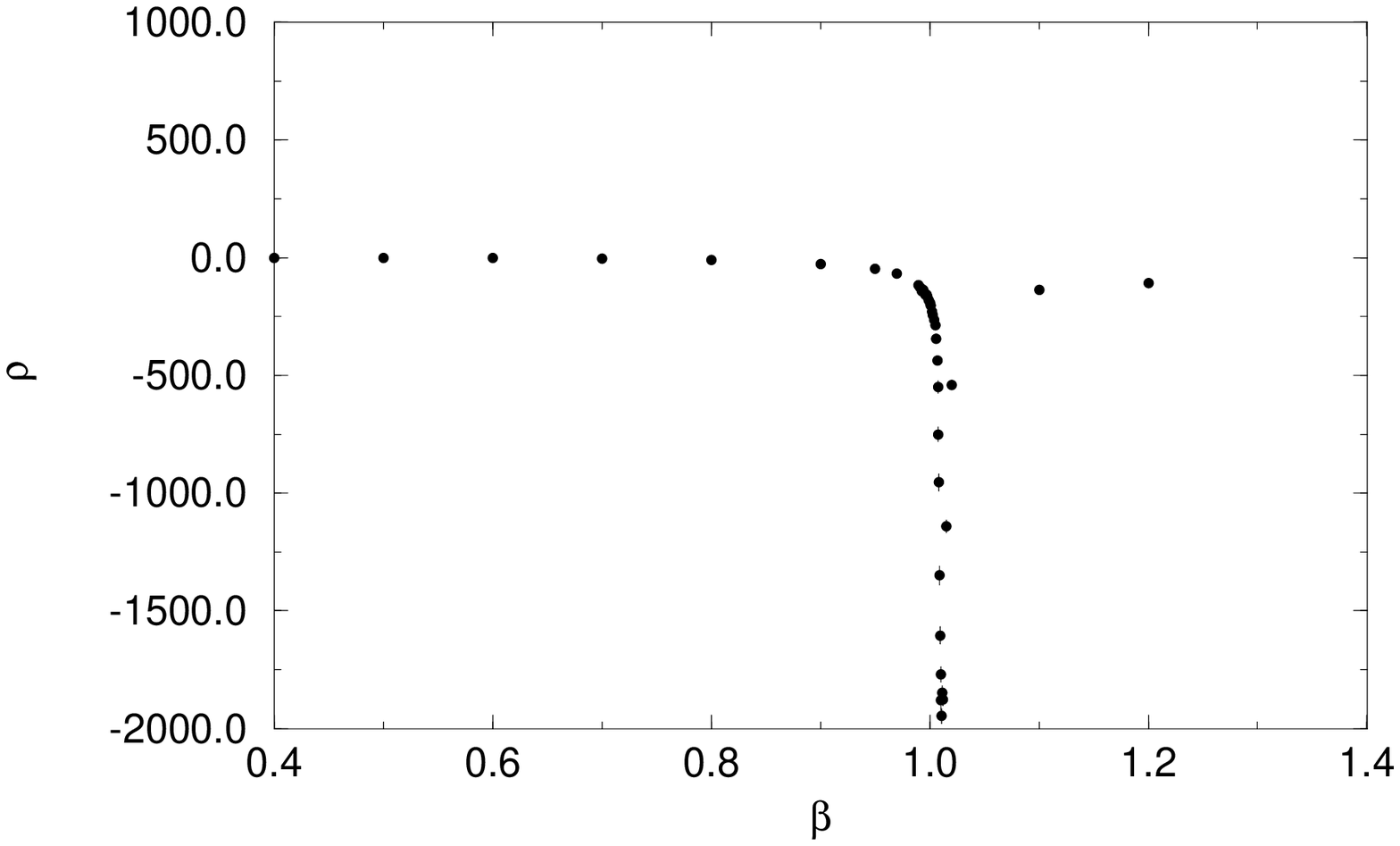}
Fig.3 $\rho_\infty$ as a function of $\beta$. 
The negative peak signals
the phase transition.
(Lattice $8^3\times16$)
\end{minipage}
\par\noindent
This can 
be appreciated from Fig.4 where a direct measurement of $\langle\mu\rangle$ 
is shown, even with large errors.

\vskip0.1in\par\noindent
\hskip0.05\textwidth
\begin{minipage}{0.90\textwidth}
\epsfxsize = 0.95\textwidth
%{\centerline{
\epsfbox{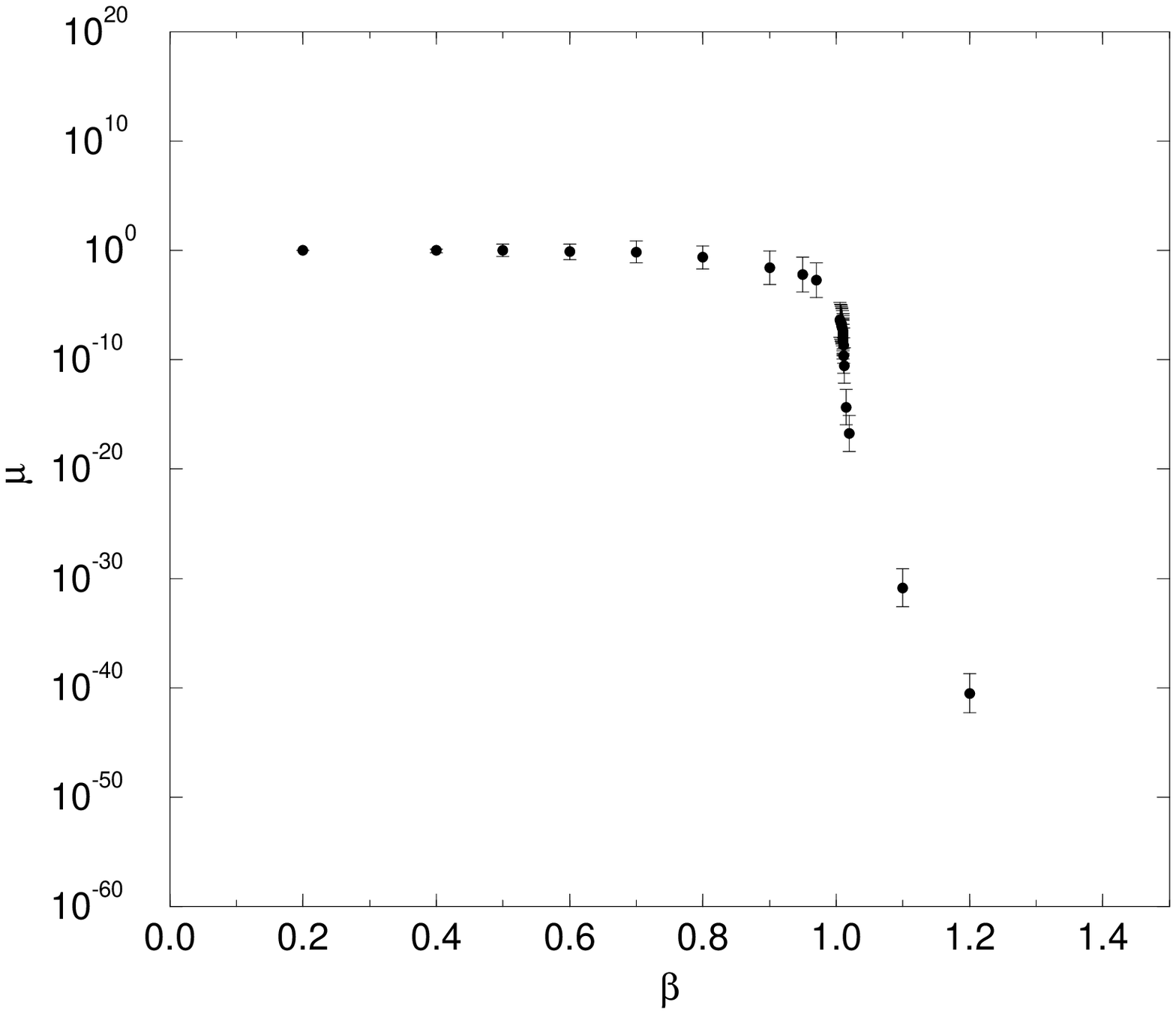}
% }}
%{\centerline{
Fig.4
$\langle\mu\rangle$ v.s. $\beta$.
Lattice $10^3\times20$.
%}}
\end{minipage}
\vskip0.1in\par\noindent
\hskip0.05\textwidth
\begin{minipage}{0.90\textwidth}
\epsfxsize = 0.95\textwidth
\epsfbox{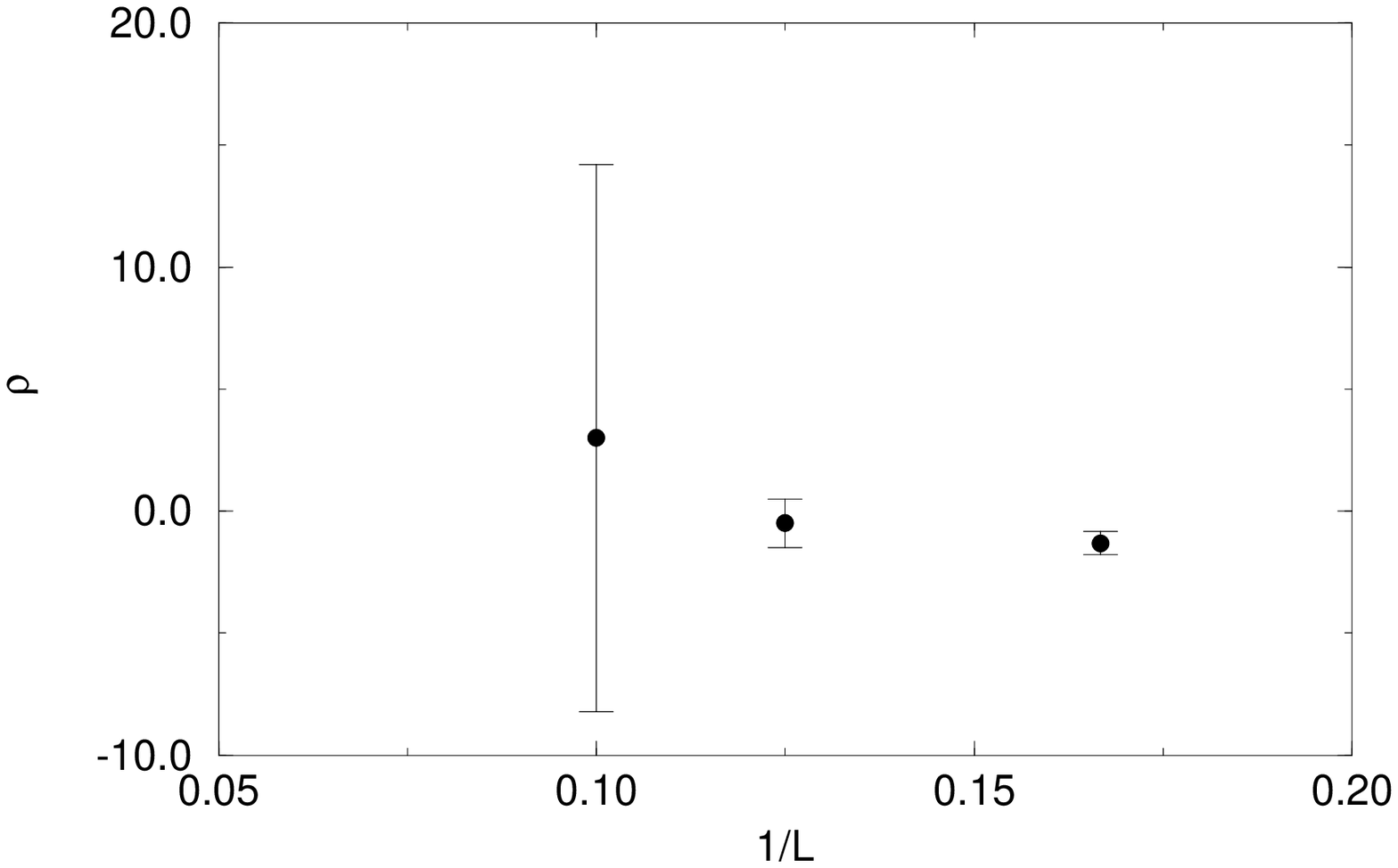}
Fig.5 $\rho_\infty$ versus $1/L$ for $\beta = 1.009$. 
\end{minipage}
\vskip0.1in\par\noindent
At $\beta > \beta_c$ the system describes 
free photons: $\mu$ and $\rho(\infty)$
can be computed in perturbation thory by a gaussian integration. The numeric
result is, for a lattice $L^3\times 2L$
\begin{equation}
\rho_\infty = -10.1\cdot L + 9.542 \qquad (\beta > \beta_c)
\label{eq:4.22}\end{equation}
$\rho_\infty$ tends to $-\infty$ in the thermodynamical limit 
$L\to \infty$, or 
$\langle\mu\rangle\to 0$. In fact $\langle\mu\rangle$ is an analytic function
of $\beta$ at finite volume, and cannot be exactly zero for $\beta > \beta_c$,
since it would be identically zero everywhere. Only as $V\to\infty$ Lee - Yang
singularities develop and $\langle\mu\rangle = 0$ for $\beta > \beta_c$
(\ref{eq:4.22}) as a respectable order parameter.

For $\beta < \beta_c$, $\rho_\infty$ tends to a finite value as $V\to\infty$.
This can be seen from fig.5 but it is also a theorem proven in ref.(16) which
generalizes the result of ref.(15) for Villain action.

\vskip0.1in
\noindent
For $\beta\sim\beta_c$ there is a phase transition, which is  weak first
order or second order. In any case the correlation length will grow as
$\beta\to\beta_c$ in a certain interval of $(\beta_c - \beta)$ with some
effective critical index $\nu$:
\begin{equation}
\xi \simeq (\beta_c - \beta)^{-\nu} \label{eq:4.23}\end{equation}
A finite size scaling analysis can be done as follow. In general, by
dimensional reasons
\begin{equation}
\mu = \mu\left(\frac{L}{\xi},\frac{a}{\xi}\right)
\label{eq:4.23a}\end{equation}
As $\beta\to\beta_c$, $a/\xi\to 0$ and $\mu \simeq\mu(L/\xi,0)$ or
\begin{equation}
\mu = f\left(L^{1/\nu}(\beta_c-\beta)\right)\label{eq:4.23b}\end{equation}
which implies in turn that
\begin{equation}
\rho = \frac{d}{d\beta}\ln\langle\mu\rangle =
- L^{1/\nu} \dfrac{ f'(L^{1/\nu}(\beta_c-\beta))}{f(L^{1/\nu}(\beta_c-\beta))}
\label{eq:4.22c}\end{equation}
or $\rho L^{-1/\nu}$ is a universal function of $L^{1/\nu}(\beta_c - \beta)$.

Data from lattices of different size will lay on the same universal curve only
for appropriate values of $\beta_c$ and $\nu$.
The best values can be then determined.
We obtain 
\begin{equation}
\beta_c = 1.01160(5)\qquad \nu = 0.29(2)\label{eq:4.22d}\end{equation}
$\mu\simeq (\beta_c-\beta)^\delta$ as $\beta\to \beta_c$, and
\[ \dfrac{\rho}{L^{1/\nu}}\simeq - \dfrac{\delta}{L^{1/\nu}(\beta_c-\beta)}
\]
For $\delta$ we obtain $\delta=1.1\pm 0.2$.

Our result is then that $\langle\mu\rangle\neq0$ for $\beta < \beta_c$ in the
thermodinamic limit, i.e. that the system is a dual superconductor for 
$\beta < \beta_c$.

We have also measured the penetration depth of the electric field, i.e. the mass
of the photon, $m$, by the method of ref.(18). $m$ properly scales  
as $\beta\to \beta_c$ with index $\nu$, 
and the indication is that it is substantially
smaller than $M$, fig.6, or that the superconductor is type II.

\par\noindent
\begin{minipage}{0.9\textwidth}
\epsfxsize = 0.85\textwidth
\epsfbox{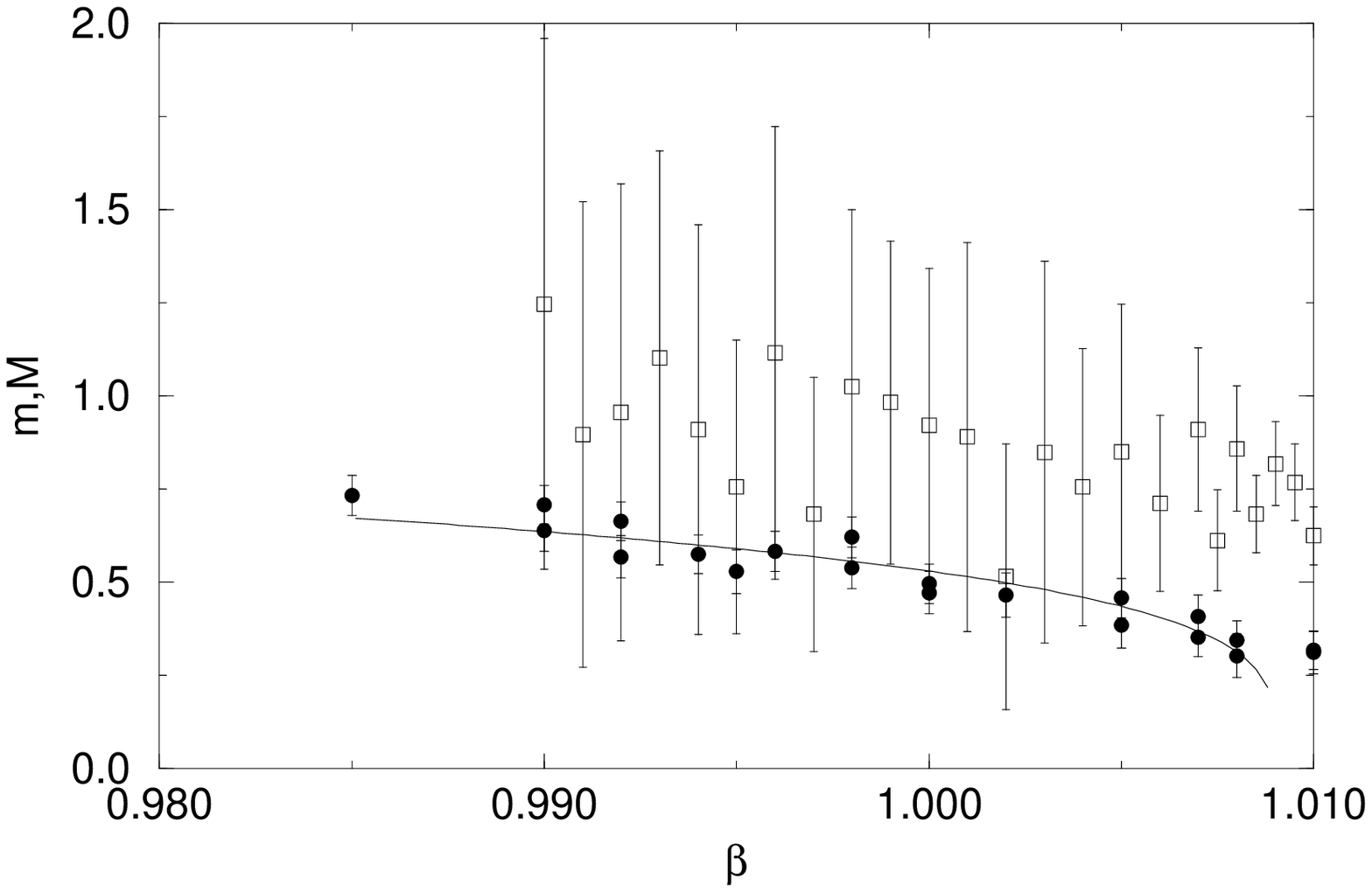}% }}
Fig.6 Mass of the monopole $M$ (squares), and mass of the 
dual photon 
$m$ (circles) vs. $\beta$.
\end{minipage}
\vskip0.1in
\noindent
An alternative method to demonstrate dual superconductivity is to detect 
London
current in the flux tube configurations between $q\bar q$\site{14}. 
For a detailed
comparison of our 
approach with ref.16
we refer to ref.13.

As for a direct determination of $\langle\mu\rangle$ it can be shown that
$\langle\mu\rangle$ has a gaussian distribution in the sense of the central
limit theorem\site{11bis}. 
However, due to the exponential dependence on $\Delta S$
the expectation value of $\langle\mu\rangle$ is not centered at the minimum of
$\Delta S$.  Let  $\Pi(y)$ be the distribution of
$y = \beta\Delta S - \beta\langle\Delta \bar S\rangle$,
the distribution probability for $\mu$ is
\[P(\mu) = \Pi\left(\ln(\frac{\mu}{\bar \mu})\right)
d \ln(\frac{\mu}{\bar\mu})\]
If $\Pi(y)$ is gaussian with width $\sigma_y$,
\[\overline{\langle\mu\rangle} = \overline\mu\exp(\langle \frac{\sigma_y^2}{2}
\rangle)\qquad \sigma_\mu = \overline\mu
\exp(\langle \frac{\sigma_y^2}{2}\rangle)\]
The fluctuations $\sigma_\mu$ are larger than $\overline{\langle\mu\rangle}$.

A careful analysis requires the account for higher cumulants. The displacement
of the maximum of $P(\mu)$ with respect to $\langle\bar\mu\rangle$ should be
kept into account when computing the so called constrained effective potential.

As a final comment it can be shown that our construction gives the same result
as that of ref.16 in the case of Villain action.

We have thus a disorder parameter which is a reliable tool to detect dual
superconductivity.

We have successfully repeated the construction for the $X-Y$ model in 3d, where
vortices condense to produce the phase transition. The result can be checked
aganst experiment (liquid $He_4$).
\section{3d $X-Y$ model$^{21}$ (liquid $He_4$)}
The model is defined on a 3d cubic lattice. An angle $\theta(i)$ is defined on
each site $i$. The action reads
\begin{equation}
S = \beta\sum_i\sum_{\mu=0}^2\left[1 - \cos(\Delta_\mu\theta(i))\right]
\label{eq:5.1}\end{equation}
$\Delta_\mu \theta(i) = \theta(i+\hat\mu) - \theta(i)$. The partition function
is
\begin{equation}
Z = \int\prod_i\dfrac{d\theta(i)}{2\pi}\exp(- S)
\label{eq:5.1a}\end{equation}
$Z$ is a periodic functional of $\theta(i)$ with period $2\pi$ (compactness).
Z is invariant under the change of variables
\begin{equation}
\theta(i) \to \theta(i) + f(i)\label{eq:5.2}\end{equation}
with arbitrary $f(i)$ and so is any correlator of compact fields, in spite of
the fact that 
the transformation
(\ref{eq:5.2}) is not a symmetry of the action. In running the
indices in \Ref{eq:5.2} from 0 to 2 we anticipate that we shall consider the
theory as the euclidean version of a 2+1 dimensional field theory.

As $\beta\to \infty$

\begin{equation}
S \simeq \frac{\beta}{2}\sum_{\mu,i}\left[\Delta_\mu(\theta_i)\right]^2
\label{eq:5.3}\end{equation}
and the theory describes a massless scalar field. At $\beta=\beta_c=0.454$ a
2nd order phase transition takes place. Below $\beta_c$ vortices are expected
to condense in the vacuum. Like for monopoles, condensation has always been
demonstrated by the drop of the density of vortices when $\beta$ raises through
$\beta_c$. We will show instead that a spontaneous symmetry breaking of the
$U(1)$ symmetry related to the conservation of vortex number takes place and
that a legitimate disorder parameter can be defined. We define
\begin{equation}
A_\mu = \partial_\mu\theta\label{eq:5.4}\end{equation}
Under the transformation \Ref{eq:5.2} $A_\mu$ undergoes a gauge transformation 
\begin{equation}
A_\mu \to A_\mu + \partial_\mu f\label{eq:5.5}\end{equation}
The invariance under \Ref{eq:5.2} means gauge invariance if the theory is
phrased in terms of $A_\mu$. From \Ref{eq:5.4} it follows
\begin{equation}
F_{\mu\nu} = \partial_\mu A_\nu - \partial_\nu A_\mu = 0
\label{eq:5.6}\end{equation}
In fact \Ref{eq:5.6} is valid apart from singularities. In terms of $A_\mu$
\begin{equation}
\theta(x) = \int_C^x\exp(\ii A_\mu d x^\mu) \label{eq:5.7}\end{equation}
and, if \Ref{eq:5.6} holds, the choice of the path $C$ used in \Ref{eq:5.7}
is irrelevant. A current $j_\mu$ can be defined as the dual of $F_{\mu\nu}$:
\begin{equation}
j_\mu = \varepsilon_{\mu\alpha\beta} \partial^\alpha A^\beta
\label{eq:5.8}\end{equation}
and
\begin{equation}
\partial^\mu j_\mu = 0 \label{eq:5.9}\end{equation}
is the analog of Bianchi identities.
The conserved quantity associated to \Ref{eq:5.9} is the vorticity
\begin{equation}
Q = \int d^2 xj^0(\vec x,t) =
\int(\vec\nabla \wedge\vec A)_0 =
\oint\vec A d\vec x\label{eq:5.10}\end{equation}
Since $\vec A=\vec\nabla\theta$, single-valuedness of the action implies that $Q
= 2\pi n$. If there are no singularities it follows from \Ref{eq:5.4} that
$\vec\nabla \wedge\vec A=0$ or, from \Ref{eq:5.10} that $Q=0$.

There exist however configuration with non trivial vorticity. An example is
\begin{equation}
\bar\theta_{(q)}(\vec x-\vec y) =
q\, {\rm arctan}\dfrac{(\vec x-\vec y)_2}{(\vec x-\vec y)_1}
\label{eq:5.11}\end{equation}
For this configuration
\begin{equation}
A^0 = 0\qquad \vec A = \frac{q}{r}\vec\nu_\theta\label{eq:5.12}\end{equation}
$\vec\nu_\theta$ being the unit vector in the  direction ($\theta$) in
polar coordinates. If $\vec A$ is the field of velocities $\bar\theta_{(q)}$
describes a vortex. For this configuration
$ j^0 = 2\pi\delta^{(2)}(\vec x-\vec y)$ and
\begin{equation}
\oint_C \vec A\cdot d\vec x = 2\pi q\label{eq:5.13}\end{equation}
if the path $C$ encloses $\vec y$ once.

A disorder parameter describing condensation of vortices can be defined. In the
continuum this is nothing but addition to the field $\theta(x)$ of any
configuration of the field $\bar\theta$ describing a vortex. The conjugate
momentum to $\theta$ being $\beta\sin\partial_0\theta = {\partial{\cal L}}/
\partial(\partial_0\theta)$
\begin{equation}
\mu(\vec y,t) = \exp\left[
\ii\beta
\int d^2 x\,\sin(\partial_0\theta(\vec x,t))\bar\theta_{(q)}(\vec x-\vec y)
\right] \label{eq:5.14}\end{equation}
analogous to \Ref{eq:4.14}.

After Wick rotation the compactified version of \Ref{eq:5.14} becomes
(see \Ref{eq:4.13})
\begin{equation}
\mu(\vec y,t) = \exp\left\{
-\beta\sum_{\vec n}\left[\cos(\Delta_0\theta(\vec n,t) -
\bar\theta_{(q)}(\vec n-\vec y)) - \cos(\Delta_0\theta(\vec n,t)\right]\right\}
\label{eq:5.15}\end{equation}
When computing correlation functions of $\mu(\vec y,t)$ \Ref{eq:5.15} produces
a change of the term in the action containing $\Delta_0\theta$ at time $t$
from $\Delta_0\theta(\vec n,t)$ to
$\Delta_0\theta(\vec n,t) - \bar\theta_{(q)}(\vec n-\vec y)$. The correlator
\begin{equation}
{\cal D}(t) = \langle \mu(\vec 0,t)\mu(\vec 0,0)\rangle
\label{eq:5.16}\end{equation}
will behave as
\begin{equation}
{\cal D}(t)\mathop\simeq_{t\to\infty} \langle\mu\rangle^2 + A \ee^{-M t}
\label{eq:5.17}\end{equation}
and, from the definition of $\mu$
\begin{equation}
{\cal D}(t) = \dfrac{Z[S+\Delta S]}{Z[S]} \label{eq:5.18}\end{equation}
The  $S + \Delta S$ 
is obtained from $S$ by the modification
\begin{eqnarray}
\Delta_0\theta(\vec n,0) &\to& \Delta_0\theta(\vec n,0) -
\bar\theta_{(q)}(\vec n-\vec y)\label{eq:5.19}\\
\Delta_0\theta(\vec n,t) &\to& \Delta_0\theta(\vec n,t) +
\bar\theta_{(q)}(\vec n-\vec y)\label{eq:5.20}
\end{eqnarray}
A change of variables $\theta(\vec n,1)\to \theta(\vec n,1) - \bar\theta_{(q)}$
reabsorbs the modification in $\Delta_0\theta(\vec n,0)$, but shifts
$\Delta_i\theta(\vec n,1) = A_i(\vec n,1)$ by a vortex
$\Delta_i\bar\theta_{(q)} = A_i^{(q)}$ and sends
\[ \Delta_0\theta(\vec n,1) \to
\Delta_0\theta(\vec n,1) - \bar\theta_{(q)}(\vec n-\vec y)\]
Similarly to what was done with $U(1)$
monopoles the construction can be repeated till
$n_0 = t$ is reached and $-\bar\theta_{(q)}$ cancels
$\bar\theta_{(q)}$ of \Ref{eq:5.20}. 

${\cal D}(t)$ describes a vortex at $\vec y = 0$ propagating in time from 0 to
$t$.
Instead of ${\cal D}(t)$ $\rho(t) = \frac{d}{d\beta}\ln {\cal D}(t)$ can be
studied. At large $t$
\begin{eqnarray}
\rho(t) &\simeq& \rho + A \ee^{-M t} \label{eq:5.21}\\
\rho &=&
2\dfrac{d}{d\beta}\ln\langle\mu\rangle\qquad
\langle\mu\rangle = exp\left[\frac{1}{2}\int_0^\beta
\rho_\infty(\beta)\,d\beta\right]
\label{eq:5.22}\end{eqnarray}

\par\noindent
\begin{minipage}{0.9\textwidth}
{\rotatebox{-90}{
\epsfxsize = 0.65\textwidth
\epsfbox{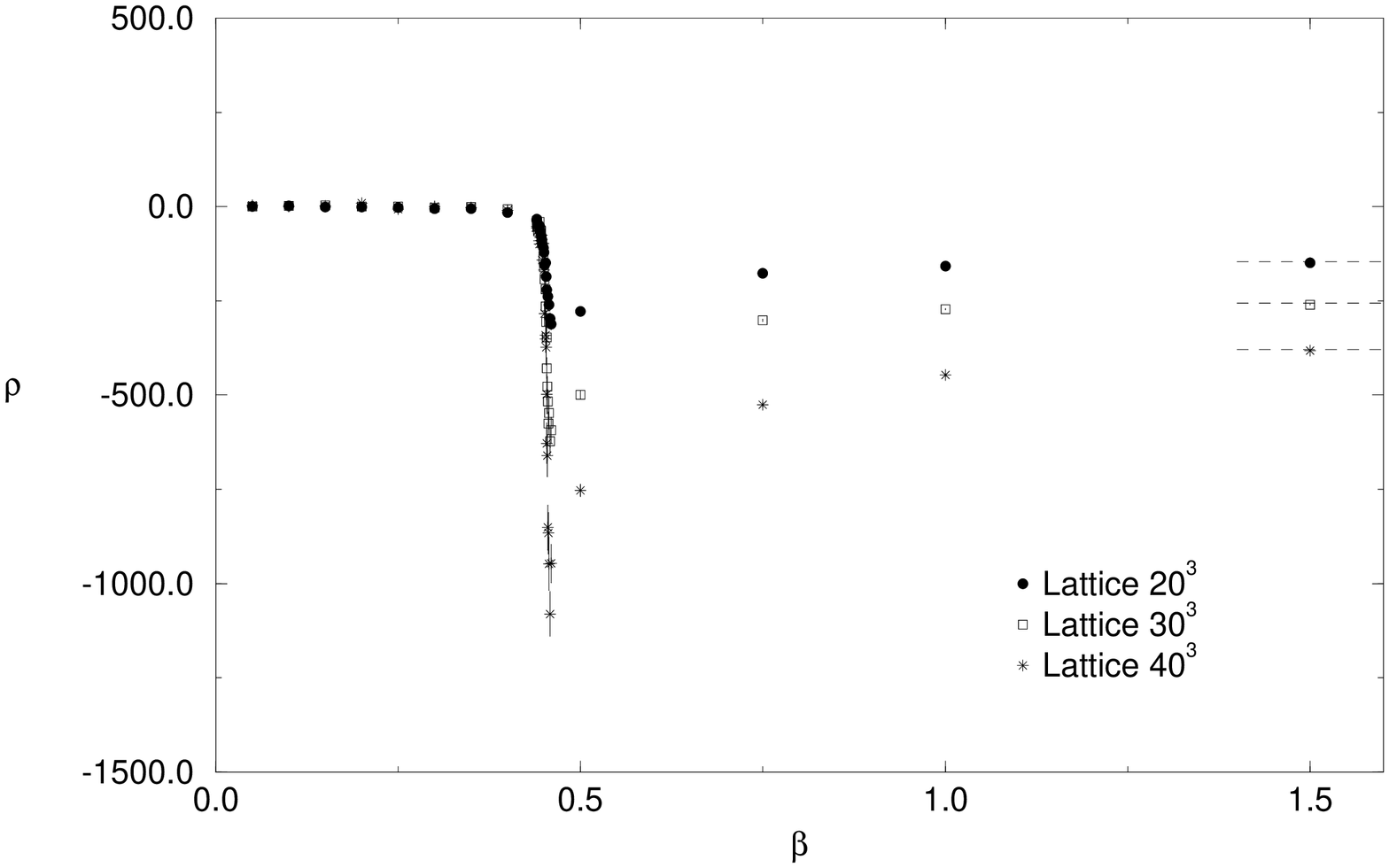}
}
}\vskip0.05in\par\noindent
Fig.7
$\rho$ vs $\beta$ for $X-Y$ model.
The peak is at the transition to superfluid.
\end{minipage}
\vskip0.2in

From \Ref{eq:5.22} the disorder parameter $\langle\mu\rangle$ can be
determined. A typical behaviour of $\rho$ is shown in fig.6. In the
thermodynamical limit
$\langle\mu\rangle\neq 0$ signals spontaneous symmetry breaking of the symmetry
\Ref{eq:5.9}, and hence condensation of vortices. At large $\beta$ the theory
is free and $\rho$ can be computed by a gaussian integration. The result for a
cubic lattice os size $L$ is
\begin{equation}
\rho = -11.33\cdot L + 72.669 \label{eq:5.24}\end{equation}
As $L\to\infty$ this implies $\langle\mu\rangle = 0$. For $\beta < \beta_c$,
$\rho$ tends to a finite value, as $L\to \infty$.

Around $\beta_c$ a finite size scaling analysis can be done 
as in the $U(1)$ model giving
\begin{eqnarray}
\nu &=& 0.669 \pm 0.065\qquad [.670(7)]\label{eq:5.25}\\
\beta_c &=& 0.4538 \pm 0.0003\qquad [.45419(2)]\label{eq:5.26}\\
\delta &=& 0.740 \pm 0.029\label{eq:5.27}\end{eqnarray}
$\delta$ is the index of $\mu$.
The numbers in parenthesis on the right are the accepted determinations 
by other methods\site{18}. The agreement is good.

\section{Monopoles in QCD. Revisiting the abelian projection$^{23}$.}
At the classical level monopoles in non abelian gauge theories can be defined
by the usual multipole expansion\site{21b,21c}.
At large distances the magnetic monopole field obeys abelian equations. By a
suitable choice of gauge the direction of the Dirac string can be chosen, and
magnetic charges are identified by a diagonal matrix in the fundamental
representation with positive or negative integer eigenvalues. For $SU(N)$, 
$N-1$
magnetic charges exist, which correspond to a group $U(1)^{N-1}$.

This classification coincides with the so called abelian projection.

Let $\vec\Phi\vec\sigma$ be any field belonging to the adjoint representation:
in what follows we shall refer to $SU(2)$ for the sake of simplicity. For
$SU(N)$ the procedure is analogous with some formal complication. A gauge
transformation $U(x)$ which diagonalizes $\vec\Phi\vec\sigma$ is called an
abelian projection. $U(x)$ will be singular at the locations where $\vec\Phi =
0$. These locations are world lines of $U(1)$ Dirac monopoles. $U(1)$ is the
residual invariance under rotation around the axis of 
$\vec\Phi\vec\sigma = \Phi\sigma_3$ after diagonalization.

These monopoles are supposed to condense and produce dual superconductivity.
There is a large arbitrariness in the choice of $\vec\Phi\vec\sigma$, i.e. in
the identification of monopoles. We will discuss this point in detail below.

The meaning of the abelian projection can be better understood\site{19} 
in the Georgi
Glashow model\site{20}. 
This is an $SO(3)$ gauge theory coupled to a triplet Higgs
\begin{equation}
{\cal L} = -\frac{1}{4}\vec G_{\mu\nu}\vec G_{\mu\nu} +
(D_\mu\vec \Phi)^\dagger(D_\mu\vec\Phi) - V(\vec \Phi)
\label{eq:6.0}\end{equation}
with
\[ D_\mu\vec \Phi = (\partial_\mu - g \vec W_\mu\wedge)\vec \Phi\]
and
\[ V(\vec \Phi) = \dfrac{\lambda}{4}\left[(\vec\Phi^2)^2 -
\mu^2\vec\Phi^2\right]\]
The model admit monopoles as soliton solutions, in the spontaneous broken
phase where $\langle\vec\Phi\rangle =
\vec\Phi^0\neq 0$.

Usually a fixed (point independent) frame of reference is used in colour space,
e.g. $\vec\xi^0_i$ $(i=1,2,3)$, the unit vectors of an orthogonal cartesian
frame. $\vec\xi^0_i\wedge \vec\xi^0_j = \varepsilon_{ijk}\vec\xi^0_k$. One can,
however, define a Body Fixed Frame (BFF) by 3 orthogonal unit vectors
$\vec\xi_i(x)$
\begin{equation}
\vec\xi_i\wedge \vec\xi_j = \varepsilon_{ijk}\vec\xi_k
\label{eq:6.1}\end{equation}
such that $\vec\xi_3(x) =\hat\Phi(x)$, is parallel to the direction $\hat\Phi(x)$
of the $\vec\Phi$ field. This system is defined up to a rotation around
$\vec\xi_3$. An element of the gauge group exists $R(x)$ such that
\begin{equation}
\vec\xi_i(x) = R^{-1}(x)\vec\xi_i^0 \label{eq:6.2}\end{equation}
By construction $R(x)$  is the gauge transformation which operates the abelian
projection, since $R(x)\vec\xi_3 = \vec\xi_3^0$. Since $\vec\xi_i^2 = 1$
\[ \partial_\mu \vec\xi_i = \vec\omega_\mu\wedge\vec\xi_i\]
or
\begin{equation}
D_\mu(\omega)\vec\xi_i \equiv (\partial_\mu - \vec\omega_\mu\wedge)\vec\xi_i = 0
\label{eq:6.3}\end{equation}
\Ref{eq:6.3} implies $[D_\mu,D_\nu] = G_{\mu\nu}(\vec \omega) = 0$. The field
$\vec\omega_\mu$ is a pure gauge. The last equality reads explicitely
\begin{equation}
\partial_\mu\vec\omega_\nu - \partial_\nu\vec\omega_\mu +
\vec\omega_\mu\wedge\vec\omega_\nu = 0
\label{eq:6.4}\end{equation}
\Ref{eq:6.4} is true apart from singularities. As a consequence of \Ref{eq:6.4}
\begin{equation}
\vec\Phi(x) = P\exp\left[\ii\int_C^x \vec\omega_\mu \cdot\vec T\,d x^\mu
\right]\vec\Phi^0
\label{eq:6.5}\end{equation}
Due to \Ref{eq:6.4} the integral in \Ref{eq:6.5} is independent of the path 
$C$.
This is true apart from singularities which can give a non trivial connection
to space time.

A t'Hooft\site{21} Polyakov\site{22} monopole 
configuration has a zero of $\vec\Phi(x)$ at the
location of the monopole, and in that point the abelian projection $R(x)$ has a
singularity. 

The singularities of $\vec\omega_\mu$ can be studied by expressing $\vec\xi(x)$
in terms of polar coordinates $\theta(x)$, $\psi(x)$, with polar axis 3 in
colour space. The singularities come from the fact that $\psi(x)$ is not
defined at the sites where $\theta(x) = 0,\pi$. In terms of $\vec
G_{\mu\nu}(\omega)$ the potentially singular term at $\theta(x) = 0,\pi$ 
is\site{19}
\begin{equation}
F^3_{\mu\nu}(x) = -\cos\theta(x)(\partial_\mu \partial_\nu -
\partial_\nu\partial_\mu)\psi(x)\label{eq:6.6}\end{equation}
The singularity exists where $\theta = 0,\pi$ or at the sites where $\hat\Phi$
is in the direction 3, and the field is parallel to $\hat\Phi$ in colour space
and abelian. The singularity is a string with flux $\pm2 n\pi$ ($n =
0,1\ldots)$. The abelian field is the field related to the residual $U(1)$
symmetry along the 3d axis. The field strength is the abelian part of $\vec
F_{\mu\nu}$, or
\begin{equation}
{\cal F}_{\mu\nu} = \hat\Phi\cdot \vec G_{\mu\nu} -
\frac{1}{g}\hat\Phi\cdot(D_\mu\hat\Phi\wedge D_\nu\hat\Phi)
\label{eq:6.7}\end{equation}
Indeed, in the abelian projected frame $D_\mu\hat\Phi = g \vec
A_\mu\wedge\vec\xi_3^0$ and
\[ \frac{1}{g}\hat\Phi(D_\mu\vec\Phi\wedge D_\nu\vec\Phi) = 
g(\vec W_\mu\wedge \vec W_\nu)\vec\Phi\]
which cancels the non abelian term of $\hat\Phi\vec G_{\mu\nu}$ in \Ref{eq:6.7}.
${\cal F}_{\mu\nu}$ is nothing but a covariant expression for the $U(1)$ 
field identified by the abelian projection.

${\cal F}_{\mu\nu}$ is a gauge invariant quantity. For a t'Hooft Polyakov
monopole configuration ${\cal F}_{\mu\nu}$ is the field of a pointlike Dirac
monopole, located at the zero of $\vec\Phi(x)$. 

In QCD there are no fundamental Higgses. The idea is that monopoles could be
defined by any composite field $\vec \Phi(x)$ in the adjoint 
representation\site{21a}.
No unique criterion is known for the choice of the operator $\vec\Phi$ in QCD.
Popular choices are\site{21a}
\begin{itemize}
\item[1)] $\hat\Phi(x)$ is the Polyakov line
\begin{equation}
P(\vec x,x_0) = \oint_{C,x_0}^{x_0} P\exp(\ii \vec A_\mu \cdot\vec T\,d x^\mu)
\label{eq:6.10}\end{equation}
The path $C$ being the line $\vec x = $~constant along the time axis closing at
infinity by periodic boundary conditions.
\item[2)] Any component $\vec F_{\mu\nu}$ of the field strength tensor.
\item[3)] The operator implicitely defined by the maximization of
\begin{equation}
\sum_{\mu,n} {\rm Tr}\left\{ \sigma_3 U_\mu(n) \sigma_3 U^\dagger_\mu(n)
\right\}\label{eq:6.11}\end{equation}
This choice is known as maximal abelian projection.
\item[4)] For $SU(3)$ $F^2_{\mu\nu}$ since, contrary to the case of $SU(2)$, it
is not a singlet, due to the $d$ algebra.
\end{itemize}
For each of these choices a $U(1)$ gauge field (actually 2 for $SU(3)$) can be
identified, which couples to monopoles, and a creation operator of monopoles
can be constructed on the same lines as for compact $U(1)$.

A strategy to answer the question whether QCD vacuum 
is a dual superconductor is
to detect the condensation of the monopoles defined by different abelian
projections in the confined phase and across the transition to the 
deconfined
phase. As shown in the previous sections a reliable tool (disorder parameter)
exists for that, which has been successfully tested in systems which are well
understood. Before proceeding to that we will test the ideas of this section on
a simple system: the Heisenberg ferromagnet.
\subsection{The Heisenberg ferromagnet$^{30}$.}
The action is
\begin{equation}
{\cal L} = \frac{1}{2}\sum_{\mu=0}^2\sum_x \Delta_\mu\vec n(x) \Delta_\mu\vec
n(x) = \sum_{\mu=0}^2\sum_x (1 - \vec n(x+\hat \mu) \vec n(x))
\label{eq:6.20}\end{equation}
$x$ runs on a cubic 3d lattice and $\vec n$ is a vector of unit length, 
$\vec n^2 = 1$.

We shall look at the model as a 2+1 dimensional field theory. The Feynman
functional
\begin{equation}
Z = \int\prod_x d\Omega(x)\,\exp(-\beta S)
\label{eq:6.21}\end{equation}
has a much bigger symmetry than the lagrangean. Any local rotation
$\vec n(x) \to R(x)\vec n(x)$ even if it does not leave ${\cal L}$ invariant,
is reabsorbed in a change of variables in the Feynman integral, leaving the
measure invariant. Assuming constant boundary condition at infinity, $\vec
n_0$, we can write
\begin{equation}
\vec n(x) = R^{-1}(x)\vec n_0 \label{eq:6.22}\end{equation}
$R^{-1}$ is determined up to a rotation along $\vec n$. As in the Georgi
Glashow model a gauge field $\vec \omega_\mu$ can be defined by introducing a
Body Fixed Frame $\vec\xi_i$, with $\vec\xi_3 = \vec n$. Then
\[ \partial_\mu\vec\xi_i = \vec\omega_\mu\wedge\vec\xi_i\]
or
\[ D_\mu\vec\xi_i = (\partial_\mu - \ii \vec\omega_\mu\vec T)\vec\xi_i = 0
\]
which implies $F_{\mu\nu}(\omega) = 0$, apart from singularities and
\begin{equation}
\vec n(x) = P\exp\left[\ii\int_{\infty,C}^x\vec\omega_\mu\cdot T\,d x^\mu
\right]\vec n_0\label{eq:6.23}\end{equation}
independent of the path $C$, $\vec \omega_\mu$ being a pure gauge.
Again this is true apart from singularities. A conserved current exists
\begin{equation}
\vec j_\mu = \frac{1}{8\pi}\varepsilon_{\mu\alpha\beta}\partial_\alpha\vec n
\wedge\partial_\beta\vec n\label{eq:6.24}\end{equation}
$\vec j_\mu$ is parallel to $\vec n$, since both $\partial_\alpha\vec n$ and 
$\partial_\beta\vec n$ are orthogonal to it. The corresponding conserved
quantity is
\begin{equation}
Q = \int d^2 x j^0(x) = \frac{1}{4\pi}
\int d^2 x \vec n\cdot(\partial_1\vec n\wedge\partial_2\vec n)
\label{eq:6.24a}\end{equation}
which is nothing but the topological charge of the 2 dimensional version of the
model. Instantons of the 2 dimensional model look as solitons of the 2+1
dimensional one.
\vskip0.1in\par\noindent
\hskip0.15\textwidth
\begin{minipage}{0.90\textwidth}
\epsfxsize = 0.7\textwidth
%{\centerline{
\epsfbox{prova.epsi}
% }}
\hskip0.1\textwidth
Fig.8
$\rho$ v.s. $\beta$. Heisenberg model.
\end{minipage}\vskip0.1in\par\noindent
$\vec j_\mu$ can be also written as
\begin{equation}
\vec j_\mu = \frac{\vec n}{8\pi}\left[\varepsilon_{\mu\alpha\beta}
\left(\vec\omega_\alpha\wedge\vec \omega_\beta
\right)\cdot\vec n\right]
\label{eq:6.25}\end{equation}
Except for the singularities corresponding to the locations of the instantons
$\vec \omega_1\wedge\vec\omega_2 = \partial_1\vec\omega_2 -
\partial_2\vec\omega_1$ and
\begin{equation}
Q = \frac{1}{4\pi}\oint\vec\omega\cdot\vec n d x
\label{eq:6.26}\end{equation}
A non trivial connection is created by the presence of solitons. A direct
calculation shows that the field $\vec\omega_\mu$ has Dirac string
singularities propagating in time from the center of the soliton.
The transition
from a magnetized phase 
to a disordered phase can be seen as a condensation of solitons.

The system undergoes a second order phase transition at 
$\beta = \beta_c\simeq 0.7$ from the
magnetized phase to a disordered phase. We have investigated if the solitons
described above condense in the disordered phase. A disorder parameter can be
constructed, on the same line as in the compact $U(1)$ and in 3d $X-Y$ model,
as the vev of the creation operator of a soliton, as follows
\begin{equation}
\langle \mu(x) \rangle = \dfrac{Z[S+\Delta S]}{Z[S]}
\label{eq:6.26a}\end{equation}
where $S+\Delta S$ is obtained from $S$ by the change
\begin{equation}
(\Delta_0\vec n(\vec x,x_0))^2 \equiv
(\vec n(\vec x,x_0+1) - \vec n(\vec x,x_0))^2\to
(R^{-1}_q\vec n(\vec x,x_0+1) - \vec n(\vec x,x_0))^2
\label{eq:6.27}\end{equation}
$R_q$ is a time independent transformation which adds a soliton of charge $q$
to a configuration.

Numerical simulations show that in the thermodinamical limit $\langle
\mu(x)\rangle$ vanishes in the ordered (magnetized) phase, is different from
zero in the disordered phase, and at the phase transition obeys a finite size
scaling law from which the critical indices and the transition temperature can
be extracted. A typical form of $\rho=
d \ln\langle\mu\rangle/d\beta$
 is shown in fig.8.

The results are still preliminary but agree with the values known from other
methods\site{24}. 
We get
\beq
\beta_c = 0.69\pm0.01\label{eq:6.28}\eeq
and
\beq \nu = 0.7\pm0.2\label{eq:6.29}\eeq
This shows that the phase transition to disorder in the Heisenberg
magnet can be viewed as condensation of solitons.

The string structure of the singularities of the field $\vec \omega_\mu$ is
similar to Georgi Glashow model, and the  field strength tensor is generated by
the topology of the field $\vec n$, even in the absence of gauge fields. Indeed
going back to the previous section, the monopoles only depend on the Higgs
field. Monopoles exposed by the abelian projection are not lattice artifacts,
but reflect the dynamics of the field $\vec \Phi$.
\section{Dual superconductivity in QCD.}
I will present the first results of a systematic
exploration
 of monopole
condensation below and above the deconfining transition in different abelian
projections\site{25,26}.

Besides the disorder parameter, we also measure, at $T=0$, the 
monopole antimonopole 
correlation function 
to extract a lower limit to the effective Higgs
mass, as well as the penetration depth of the electric field, i.e. the mass of
the photon which produces (dual) Meissner effect. We use for that APE QUADRIX
machines. 

The creation operator for a monopole is constructed in analogy with the $U(1)$
operator as follows: the $\Pi^{0i}$ plaquettes in the action at the time, say
$n_0 = 0$ when the monopole is created are modified as follows
\begin{eqnarray}
\Pi^{0i}(\vec n,n_0) &=&
{\rm Tr}\left\{ U_i(\vec n,0) U_0(\vec n + \hat i,0) U^\dagger_i(\vec n,1)
U_0^\dagger(\vec n,0)\right\}\nonumber\\
&\to&
{\rm Tr}\left\{ U'_i(\vec n,0) U_0(\vec n + \hat i,0) U^\dagger_i(\vec n,1)
U_0^\dagger(\vec n,0)\right\}\label{eq:7.1}\end{eqnarray}
In \Ref{eq:7.1}
\begin{equation}
U'(\vec n,0) =
\ee^{\ii \Lambda(n)}\exp(\ii \hat\Phi\vec\sigma \frac{b_i^\bot(n)}{2})
U_i(\vec n,0) \exp(\ii \hat\Phi\vec\sigma \frac{b_i^\bot(n)}{2})
\ee^{-\ii \Lambda(n+1)}
\label{eq:7.3}\end{equation}
We call $S + \Delta S$ the resulting action. Then
\begin{equation}
\langle\mu\rangle = \dfrac{Z[S+\Delta S]}{Z[S]}\label{eq:7.2}\end{equation}
The vector potential describing the field of the monopole $b_i$ has been split
in a transverse part $b_i^\bot$ with $\partial_i b_i = 0$ and a gauge
$\partial_i\Lambda$.
The gauge dependence in \Ref{eq:7.2} can be reabsorbed in a redefinition of the
temporal links, which leaves the measure invariant.

A redefinition of $U_i$ to $\tilde U_i =
\ee^{\ii\sigma_3\frac{b_i^\bot(n)}{2}} U_i
\ee^{\ii\sigma_3\frac{b_i^\bot(n)}{2}}$ 
in the abelian projected gauge can be reabsorbed in a change of variables which
leaves the measure invariant. The space plaquettes however acquire in each 
link factors 
\beq
\ee^{\ii\vec\sigma\hat\Phi\frac{b_i^\bot(n)}{2}} \tilde U_i
\ee^{\ii\vec\sigma\hat\Phi\frac{b_i^\bot(n)}{2}}
\label{eq:7.4}\eeq

In the abelian projected gauge the generic $\tilde U_i$ canbe written as
\beq
\tilde U_i =
\ee^{\ii\sigma_3 \alpha_i}\ee^{\ii\sigma_2\gamma_i}
\ee^{\ii\sigma_3 \beta_i}
=
\ee^{\ii\sigma_3 \alpha_i}\ee^{\ii\sigma_2\gamma_i}
\ee^{-\ii\sigma_3 \alpha_i}\ee^{\ii\sigma_3 (\alpha_i+\beta_i)}
= V \ee^{\ii\sigma_3 \theta_i}
\label{eq:7.5}\eeq
$ \theta_i = \alpha_i + \beta_i$. $\exp(\ii \sigma_3\theta_i)$
is the abelian link. The transformation \Ref{eq:7.3} adds $b_i$ to $\theta_i$
 or adds the magnetic field of the monopole to the abelian 
magnetic field
\beq
\Delta_i\theta_j - \Delta_j\theta_i \to
\Delta_i\theta_j - \Delta_j\theta_i + 
\Delta_i b_j - \Delta_j b_i \label{eq:7.6}\eeq
An abelian projected monopole has been created.
$\langle\mu\rangle$ or 
better $\rho = d (\ln\langle\mu\rangle)/d\beta$ is measured across the 
deconfining phase transition. Typical behaviours are shown in fig.9 and 
fig.10 for the
monopole defined by the Polyakov loop.
\vskip0.1in
\par\noindent
\hskip0.25\textwidth
\begin{minipage}{0.50\textwidth}
\epsfxsize = 0.90\textwidth
%{\centerline{
\epsfbox{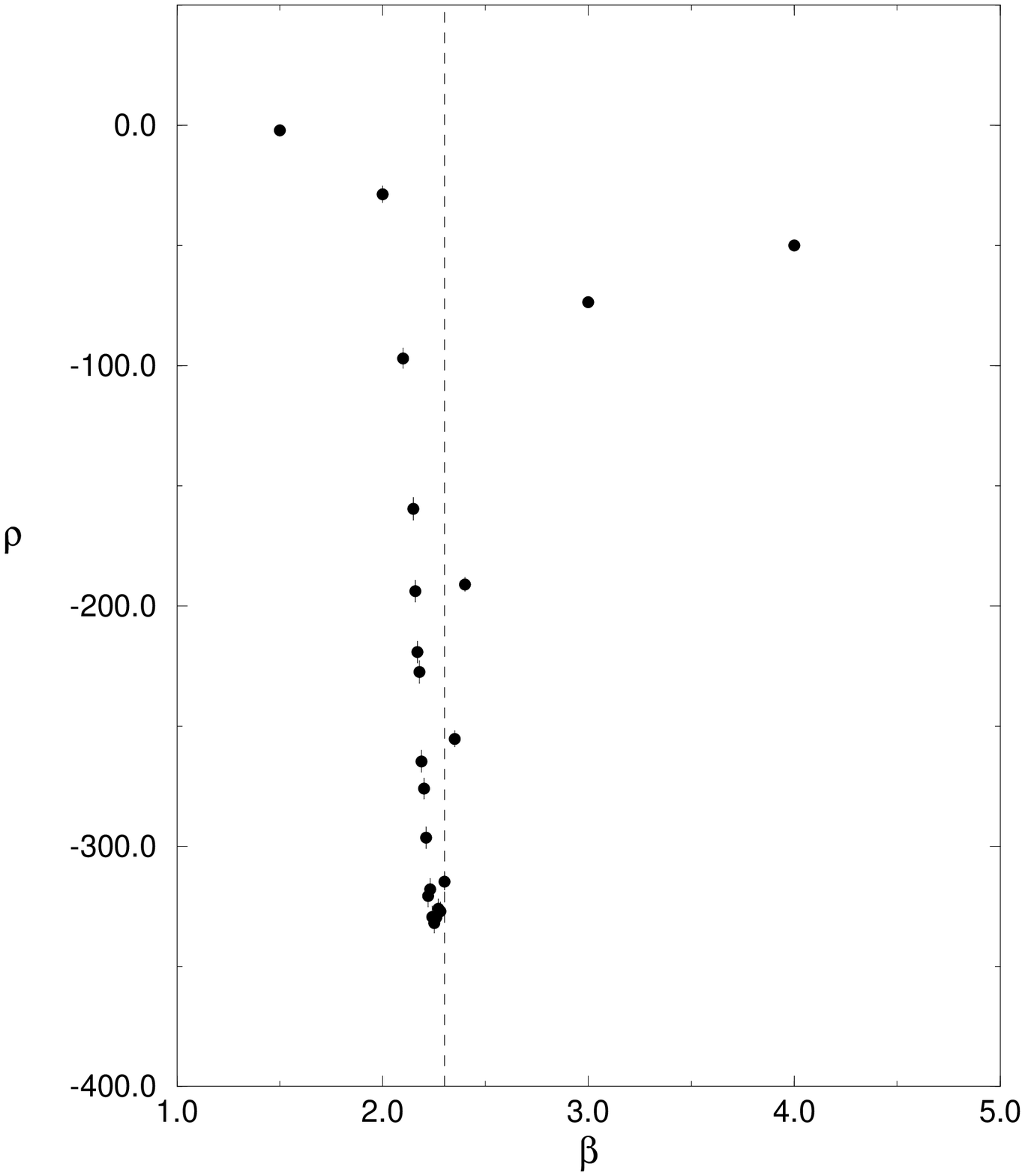}
% }}
%{\centerline{
Fig.9 $\rho$ v.s. $\beta$ $SU(2)$ gauge theory.
(Lattice $12^3\times4$)
%}}
\end{minipage}
\vskip0.1in\par\noindent
\hskip0.05\textwidth
\begin{minipage}{0.90\textwidth}
\epsfxsize = 0.95\textwidth
%{\centerline{
\epsfbox{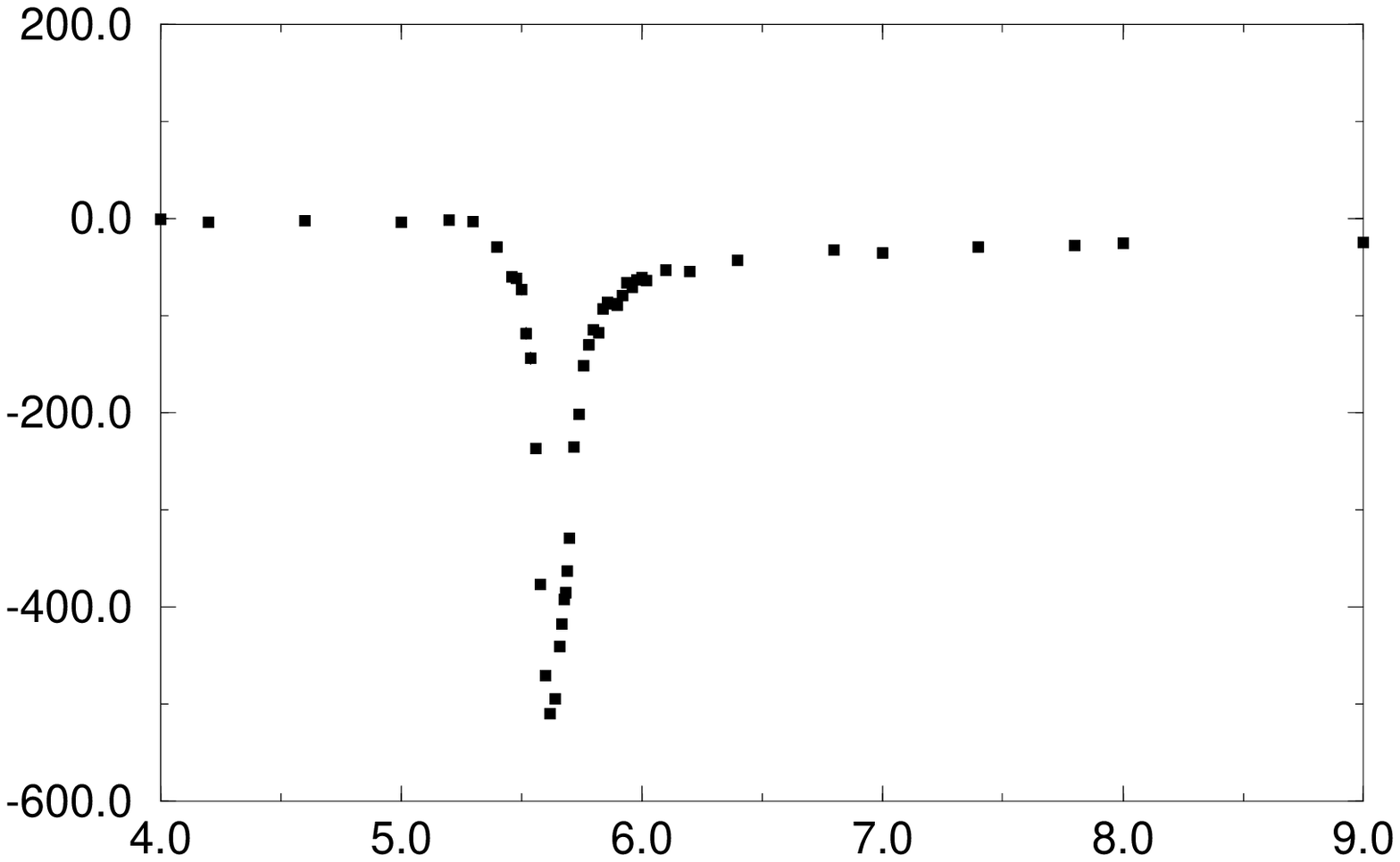}
% }}
%{\centerline{
Fig.10 $\rho$ v.s. $\beta$ $SU(3)$ gauge theory.
(Lattice $12^3\times4$)
%}}
\end{minipage}
\vskip0.1in\par\noindent
A careful analysis of the thermodynamical limit produces evidence of dual superconductivity below the deconfining temperature $T_c$, for different choices of
 $\hat\Phi$,Polyakov line, $F_{\mu\nu}$, max abelian.

We need more work and more statistics to determine the critical indices and 
the type of superconductor.

\section{Discussion}
Most of the success of the dual superconductivity mechanism for confinement is
presently based  on the abelian dominance and on the monopole dominance 
numerically observed in the maximal abelian projection\site{27,28,29}. 
In $SU(2)$ gauge theory after max abelian projection, which amounts to maximize the quantity
\[ A = \sum_{n,\mu}{\rm Tr}\left\{
\sigma_3 U_\mu(n) \sigma_3 U^\dagger_\mu(n)\right\} \]
with respect to gauge transformation, all the links are on the average 
parallel to 3 axis within $80 - 90\%$. 
The string tension computed from the abelian part of the Wilson loops accounts 
for $80 - 90\%$ of the full string tension. Similar behaviour is found for many
other quantities. 
This is called abelian dominance.
In addition the abelian plaquettes can be split as in 
\Ref{eq:4.6} in a monopole part, $n_{\mu\nu}$ and the residual angle 
$\bar\theta_{\mu\nu}$, usually called Coulomb field. The empirical observation
 is again that the contribution of the Coulomb part to the abelian quantities 
is a small factor, so that, for example, the string tension is dominated 
in fact by the contribution of the abelian monopoles (monopole dominance).

Apparently such dominance is not observed in other abelian 
projections, except maybe in the Polyakov line projection, where, 
if the string tension is measured 
from the correlation between Polyakov lines it is $100\%$ abelian 
dominated by construction.

The more or less explicitely expressed idea is then that some abelian 
projection (the max abelian) is better than others and is `
`the abelian projection'', identifying the degrees of freedom relevant for 
confinement. 
On the one hand
the fact that in this projection links are in the abelian direction at
 $80 - 90\%$ 
implies dominance as a kinematical fact: on the 
other hand maybe it is non trivial that such a gauge exists.

The attitude presented in this paper is somewhat complementary: dual 
superconductivity is related to symmetry, and the way to detect it is to 
look for symmetry.
From this point of view, independent of possible abelian dominances, what we 
find is more similar to the idea of t'Hooft that all abelian projections
are physically equivalent. We find  condensation of max abelian, 
Polyakov line, field strength monopoles.

There are few aspects of the mechanism which need further understanding.

\begin{itemize}
\item[1)] Any abelian projection 
implies that the gluons corresponding to the residual $U(1)$'s
have no electric charge with respect to them, and hence are not 
confined. This contrasts with the observation reported in sect. 1, 
about the string tension in the adjoint representation.
\item[2)] If the mechanism of confinement were superconductivity 
produced by condensation of the monopoles defined by some abelian 
projection then the electromagnetic field in the flux tubes 
joining  $q-\bar q$ pairs should belong to the projected $U(1)$. Lattice 
exploration show that it is isotropically distributed in color 
space\site{5a,30}.
\item[3)] There are infinitely many abelian projections which can be 
obtained from each other by continuous transformations, e.g. by 
shifting the zero of
$\vec\Phi(x)$. If  one of them were privileged, it is hard to understand 
how others, 
which differs by small continuous changes, could be so different.
\end{itemize}
Most problably a more complicated and new mechanism is at work, a kind of non 
abelian dual superconductivity, which manifests itself as abelian 
superconductivity  in different
abelian projected gauges. Our aim is to understand it.

It is a special pleasure to thank all the 
collaborators who contributed to this research, in particular 
Giampiero Paffuti, Manu Mathur and Luigi Del Debbio.

{\referencestyle

}

\end{document}